\documentclass{article}

\usepackage{PRIMEarxiv}

\usepackage[utf8]{inputenc}
\usepackage[T1]{fontenc}
\usepackage[hidelinks]{hyperref}
\usepackage{url}
\usepackage{booktabs}
\usepackage{amsfonts}
\usepackage{amsmath}
\usepackage{nicefrac}
\usepackage{microtype}
\usepackage{fancyhdr}
\usepackage{graphicx}
\usepackage{pdfpages}
\usepackage[numbers,sort&compress]{natbib}

\title{Ending the NIH embargo accelerated public access to funded research, but substantial delays remain}

\author{
  Haining Wang\thanks{Correspondence: \texttt{hw56@iu.edu}; 410 W 10th St, Suite 3000, Indianapolis, IN 46202. Classification: Social Sciences / Social Sciences. Preprint: deposited on arXiv under a CC BY 4.0 license.} \\
  Department of Biostatistics and Health Data Science, Indiana University School of Medicine
}

\makeatletter
\let\arxiv@paragraph\paragraph
\renewcommand{\paragraph}{\@ifstar{\arxiv@paragraphstar}{\arxiv@paragraphplain}}
\newcommand{\arxiv@paragraphstar}[1]{\arxiv@paragraph*{#1.}}
\newcommand{\arxiv@paragraphplain}[1]{\arxiv@paragraph{#1.}}
\makeatother
\newcommand{\figwidth}{\textwidth}
\begin{document}
\maketitle
\thispagestyle{plain}

\begin{abstract}
NIH-funded articles accepted from 1 July 2025 must be public in PubMed Central (PMC) on the day of publication, ending a 12-month embargo in place since 2008. Whether ending the embargo has made articles available sooner has not been measured. Using 1,880 NIH-funded articles published January to April 2026 and 1,880 from 2025, each matched to a same-journal, same-month article that PubMed does not list as US government supported, we compare the change in availability, dated by the day each article went live in PMC. By four months, availability had risen 19.4 percentage points more among NIH-funded articles than among comparison articles. Within a week of publication, however, 6 to 8\% of articles in every group and year were available, only 0.1\% of NIH-funded articles on the day itself, and the policy effect was 0.3 percentage points. Articles reach PMC by two routes: publishers deposit the final version, or the accepted manuscript is deposited with the author's approval. Two thirds of the gain came through author manuscripts, whose median wait from publication, among those live by the data cut, fell from 374 days in 2025 to 70 in 2026; final versions from publishers moved from 44 days to 36. Agencies ending their embargoes should count days to public access, not only eventual deposit, and test publisher-triggered submission at acceptance.

\end{abstract}


Since 2008, NIH has required deposit of its funded articles in PubMed Central (PMC) but let publishers delay public release for up to 12 months \cite{nih2008}. NIH announced the end of the embargo in December 2024 \cite{nih2024} and advanced it to articles accepted from 1 July 2025 \cite{nih2025}. Ending the embargo raised the share of NIH-funded articles free to read at four months by about one in five, but few by the day of publication: the policy asks for the manuscript at acceptance \cite{nih2024}, and most manuscripts were deposited weeks after publication.

We compared NIH-funded journal articles in PubMed published January to April 2025 and 2026 (SI Appendix, sections S1 and S2). Publication year stands in for policy exposure: among NIH-funded articles with a PMC accepted date (all final published versions), 98.0\% of the 2026 cohort were accepted on or after 1 July 2025, against 1.5\% of the 2025 cohort (SI Appendix, section S1). Each NIH-funded article was matched to a same-journal, same-month comparison article that PubMed did not index as US government supported. The outcome is being live in PMC within a given number of days of publication, dated by the day the article went live (SI Appendix, section S3). The estimate is a difference-in-differences \cite{angrist2009}: the 2025-to-2026 change among NIH-funded articles minus that among comparison articles in the same journal-month cells (SI Appendix, section S4). It assumes that, without the policy, availability would have changed alike in both groups, which the 2024-to-2025 contrast checks.

\section*{Results}
\subsection*{One article in five more by four months, but no detectable change within a week}
Across 327 journals and 638 journal-month cells, the difference-in-differences at 120 days was 19.4 percentage points (95\% CI 14.3 to 25.2; Fig.~\ref{fig:main}\textit{A}), when 67.8\% of 2026 NIH-funded articles were in PMC against 41.7\% of comparison articles. At 7 days the estimate was 0.3 points ($-$1.7 to 1.8), and only 0.1\% of 2026 NIH-funded articles were in PMC by the publication date. At 30 days it was 7.1 points (4.0 to 10.4). The gain arrived weeks to months after publication, and at four months a third of NIH-funded articles were still not in PMC. Across all NIH-funded articles in the matching pool, matched or not, 9.2\% (2025) and 6.2\% (2026) were in PMC within 7 days, and 43.5\% and 63.0\% within 120 days (SI Appendix, section S1). Before the policy, the 2024-to-2025 contrast on the 277 journals shared with the main analysis was 1.4 points at 120 days and 0.5 at 7 days, with intervals excluding large changes either way (Fig.~\ref{fig:main}\textit{A}; SI Appendix, section S2). The gain came from journals that are not fully open access. In the 266 journals not listed in the Directory of Open Access Journals (DOAJ, which lists only journals whose every article is free to read on publication), the share of NIH-funded articles in PMC by 120 days rose from 24\% in 2025 to 55\% in 2026, against 16\% to 19\% for comparison articles (Fig.~\ref{fig:main}\textit{B}), and the estimate was 28.0 points (22.5 to 33.4). In the 61 DOAJ journals, where most articles were already in PMC (88 to 94\% in every group and year by 120 days), it was 2.1 points ($-$1.0 to 6.8; both in Fig.~\ref{fig:main}\textit{A}). Of the 327 journals, 225 have no PMC deposit agreement, and there the estimate was 27.4 points (20.2 to 34.7), against 21.5 (0.4 to 42.1) in 28 journals that deposit NIH-funded articles only and 3.3 ($-$0.9 to 8.7) in 74 that deposit all articles (SI Appendix, section S5).

\subsection*{Author manuscripts carried most of the gain, and most were deposited after publication}
\begin{figure*}[t]
\centering
\includegraphics[width=\figwidth]{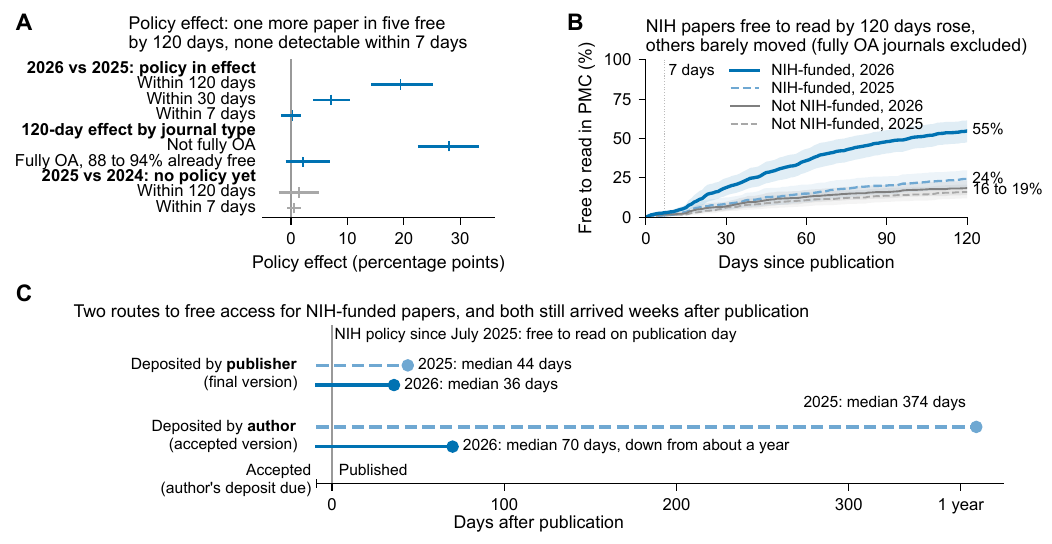}
\caption{\textbf{The 2025 NIH public-access policy and availability in PubMed Central.} (\textit{A}) Difference-in-differences in percentage points with 95\% confidence intervals: by horizon across all 327 journals; at 120 days in the 266 journals outside the Directory of Open Access Journals (DOAJ), which lists fully open-access (OA) journals, and in the 61 DOAJ journals ($N$ = 623); and the 2024-to-2025 contrast one year earlier. (\textit{B}) Share free to read (live) in PMC by day after publication, for NIH-funded and comparison (not NIH-funded) articles published January to April 2025 and 2026, in the 266 journals outside DOAJ ($N$ = 1,257 per group and year; 1,255 for NIH-funded, 2026). Bands are 95\% journal-cluster bootstrap confidence intervals; labels are shares at 120 days. (\textit{C}) The two routes into PMC, before (2025) and after (2026) the policy, lines run from median acceptance, when NIH asks for the author manuscript, to median go-live, in days after publication, among NIH-funded articles live by the data cut: final published versions deposited by journals and publishers ($N$ = 1,102 and 1,141 deposited; 1,101 and 1,107 live) and author manuscripts, approved by the author in the NIH Manuscript Submission system (NIHMS; $N$ = 618 and 472 deposited; 616 and 436 live), which publishers may initiate.}
\label{fig:main}
\end{figure*}

Author manuscripts, which reach PMC through the NIH Manuscript Submission system (NIHMS), carried 13.0 of the 19.4 points at 120 days, and final published versions carried 6.5. For manuscripts live by the data cut, the median wait from publication fell from 374 days in 2025 to 70 in 2026; for final published versions it moved from 44 to 36 (Fig.~\ref{fig:main}\textit{C}). Of the 295 manuscripts with a PubMed accepted date on or after 1 July 2025, only 2.7\% were submitted within a week of acceptance. The median submission came 72 days after acceptance, while publication followed acceptance by a median 9 days. Of all 472 NIH-funded 2026 manuscripts deposited in PMC, 93\% were submitted after publication. The 436 that had gone live did so a median 13 days after submission. Among manuscripts already live, most of the observed delay preceded submission to NIHMS.

\section*{Discussion}
Ending the embargo shortened the wait for author manuscripts that went live from about a year to about ten weeks (Fig.~\ref{fig:main}\textit{C}), and most of the remaining delay came before deposit. NIH asks for the manuscript at acceptance \cite{nih2024}, but NIHMS learns of an article only when an author or publisher submits it, and NIH checks compliance at progress reports \cite{nih2012}, which ask whether a deposit was made. Manuscripts from publishers that had volunteered to start the deposit were submitted after publication about as often as the rest (90.7\% against 95.6\%; SI Appendix, section S5), so volunteering alone did not bring deposit forward. For every agency now ending its embargo under the 2022 White House directive \cite{nelson2022}, two steps follow: count days to public access, not only eventual compliance, and test publisher-triggered submission at acceptance with release timed to publication, so that access depends less on author action. Whether that reaches publication day is untested here. The pattern matches earlier evidence that compliance with funder access mandates follows the funder's infrastructure and enforcement more than the discipline, and is lower where the mandate allows deposit after publication \cite{lariviere2018}. An embargo governs how long publishers may withhold an article. The deposit step, and PMC processing after it, governs when the public can read it, and with the embargo gone, that step sets the pace.

\section*{Materials and Methods}
NIH-funded articles were identified in PubMed by the publication types ``Research Support, N.I.H., Extramural'' or ``Research Support, N.I.H., Intramural'' with ``Journal Article'' for January to April of 2024, 2025, and 2026 (47,176 unique PMIDs). Lags run from the earlier of the PubMed and OpenAlex publication dates (SI Appendix, sections S1 and S3). Journal identity and DOAJ status are from the OpenAlex snapshot of 26 June 2026 \cite{priem2022}. Comparison candidates were OpenAlex articles with a PMID in the same journal, year, and month that PubMed did not index as US government supported or with an NIH or NSF grant agency. Articles without a PMC identifier on the list of 12 September 2026 count as not available. PMC's OAI-PMH service gives the go-live and release dates, the author-manuscript flag and, for manuscripts, the NIHMS submission date. Accepted dates are from PMC and, for manuscripts, PubMed. Cells are weighted by their matched NIH-funded articles, and confidence intervals are percentile intervals from 2,000 journal-cluster bootstrap draws.

\paragraph*{AI-assisted editing} Claude Fable 5.1 was used for copyediting. The author reviewed all content and takes responsibility for the final manuscript.

\paragraph{Data, materials, and software availability.}
Code, article-level data, and every result table are at \url{https://github.com/Wang-Haining/nih_embargo}, archived at Zenodo \cite{wang2026code} (SI Appendix, section S7).

\paragraph{Author contributions.} H.W. designed the study, performed the analysis, and wrote the paper.

\paragraph{Competing interests.} The author declares no competing interest.

\begingroup
\raggedright
\renewcommand{\bibfont}{\small}
\bibliographystyle{unsrtnat}
\bibliography{content/references}

\begin{thebibliography}{9}
\providecommand{\natexlab}[1]{#1}
\providecommand{\url}[1]{\texttt{#1}}
\expandafter\ifx\csname urlstyle\endcsname\relax
  \providecommand{\doi}[1]{doi: #1}\else
  \providecommand{\doi}{doi: \begingroup \urlstyle{rm}\Url}\fi

\bibitem[{National Institutes of Health}(2008)]{nih2008}
{National Institutes of Health}.
\newblock Revised policy on enhancing public access to archived publications
  resulting from {NIH}-funded research.
\newblock Notice NOT-OD-08-033, 2008.
\newblock Available at
  \url{https://grants.nih.gov/grants/guide/notice-files/NOT-OD-08-033.html}.
  Accessed 13 September 2026.

\bibitem[{National Institutes of Health}(2024)]{nih2024}
{National Institutes of Health}.
\newblock 2024 {NIH} public access policy.
\newblock Notice NOT-OD-25-047, 2024.
\newblock Available at
  \url{https://grants.nih.gov/grants/guide/notice-files/NOT-OD-25-047.html}.
  Accessed 13 September 2026.

\bibitem[{National Institutes of Health}(2025)]{nih2025}
{National Institutes of Health}.
\newblock Revision: Notice of updated effective date for the 2024 {NIH} public
  access policy.
\newblock Notice NOT-OD-25-101, 2025.
\newblock Available at
  \url{https://grants.nih.gov/grants/guide/notice-files/NOT-OD-25-101.html}.
  Accessed 13 September 2026.

\bibitem[Angrist and Pischke(2009)]{angrist2009}
Joshua~D. Angrist and J{\"o}rn-Steffen Pischke.
\newblock \emph{Mostly Harmless Econometrics: An Empiricist's Companion}.
\newblock Princeton University Press, Princeton, NJ, 2009.
\newblock ISBN 978-0-691-12035-5.

\bibitem[{National Institutes of Health}(2012)]{nih2012}
{National Institutes of Health}.
\newblock Upcoming changes to public access policy reporting requirements and
  related {NIH} efforts to enhance compliance.
\newblock Notice NOT-OD-12-160, 2012.
\newblock Available at
  \url{https://grants.nih.gov/grants/guide/notice-files/NOT-OD-12-160.html}.
  Accessed 16 October 2026.

\bibitem[Nelson(2022)]{nelson2022}
Alondra Nelson.
\newblock Ensuring free, immediate, and equitable access to federally funded
  research.
\newblock Memorandum, Office of Science and Technology Policy, Executive Office
  of the President, 2022.
\newblock Available at
  \url{https://bidenwhitehouse.archives.gov/wp-content/uploads/2022/08/08-2022-OSTP-Public-Access-Memo.pdf}.
  Accessed 13 September 2026.

\bibitem[Larivi{\`e}re and Sugimoto(2018)]{lariviere2018}
Vincent Larivi{\`e}re and Cassidy~R. Sugimoto.
\newblock Do authors comply when funders enforce open access to research?
\newblock \emph{Nature}, 562:\penalty0 483--486, 2018.
\newblock \doi{10.1038/d41586-018-07101-w}.

\bibitem[Priem et~al.(2022)Priem, Piwowar, and Orr]{priem2022}
Jason Priem, Heather Piwowar, and Richard Orr.
\newblock {OpenAlex}: A fully-open index of scholarly works, authors, venues,
  institutions, and concepts.
\newblock arXiv [Preprint], 2022.
\newblock Available at \url{https://arxiv.org/abs/2205.01833}. Accessed 13
  September 2026.

\bibitem[Wang(2026)]{wang2026code}
Haining Wang.
\newblock nih\_embargo: code and data for a study of the 2025 {NIH} public
  access policy.
\newblock Zenodo, 2026.
\newblock Version v1.0.1. \url{https://doi.org/10.5281/zenodo.23826370}.
  Deposited 30 October 2026.

\end{thebibliography}
\endgroup

\clearpage
\includepdf[pages=-]{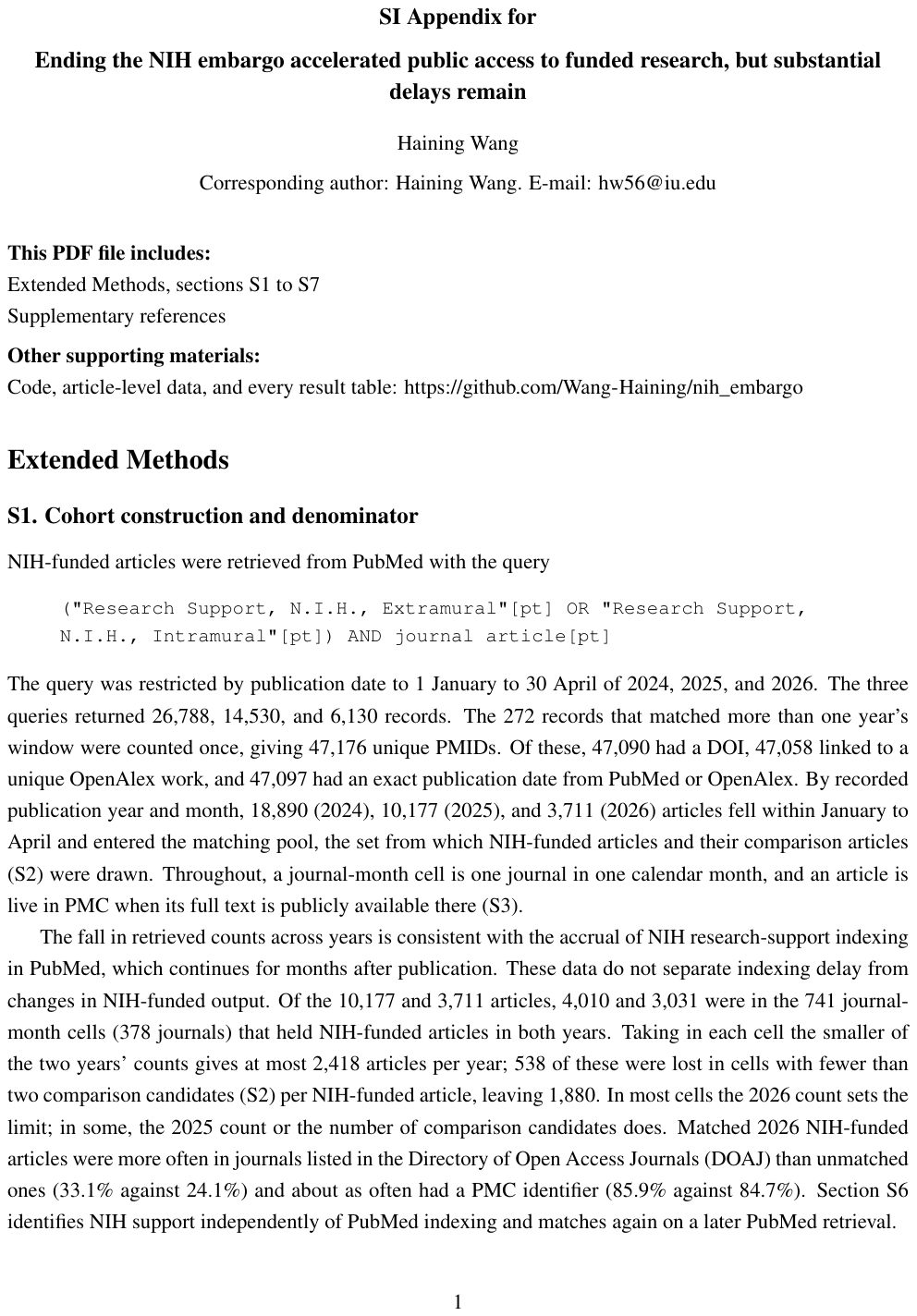}

\end{document}